\documentclass[jkps,a4paper,prd,longbibliography,twocolumn,superscriptaddress]{revtex4-1}
\usepackage{amssymb,amsmath,amsfonts,graphicx,color,xcolor,multirow,ulem,kotex}

\begin{document}
\title[]{Critical behaviors of cascading dynamics on multiplex two-dimensional lattices}
\author{Jeehye \surname{Choi}}
\affiliation{Research Institute for Nanoscale Science and Technology, Chungbuk National University, Cheongju, Chungbuk 28644, Korea}
\author{Byungjoon \surname{Min}}
\thanks{bmin@cbnu.ac.kr}
\affiliation{Research Institute for Nanoscale Science and Technology, Chungbuk National University, Cheongju, Chungbuk 28644, Korea}
\affiliation{Department of Physics, Chungbuk National University, Cheongju, Chungbuk 28644, Korea}
\author{K.-I. \surname{Goh}}
\thanks{kgoh@korea.ac.kr}
\affiliation{Department of Physics, Korea University, Seoul 02841, Korea}

\date{\today}

\begin{abstract}
We study the critical phenomena of viable clusters in multiplex two-dimensional lattices
using numerical simulations. We identify viable sites on multiplex lattices using two  
cascading algorithms: the cascade of activations ($\mathrm{CA}$) and deactivations ($\mathrm{CD}$). 
We found that the giant viable clusters identified by $\mathrm{CA}$ and $\mathrm{CD}$ processes 
exhibit different critical behaviors. 
Specifically, the critical phenomena of $\mathrm{CA}$ processes are consistent with 
the ordinary bond percolation on a single layer but 
$\mathrm{CD}$ processes exhibit the critical behaviors consistent with 
mutual percolation on multiplex lattices. In addition, we computed the susceptibility of 
cascading dynamics by using the concept of ghost field. Our results suggest that the 
$\mathrm{CA}$ and $\mathrm{CD}$ processes generate viable clusters in different ways. 
\end{abstract}
\maketitle

\section{Introduction}

The elements of a complex system often function properly only if multiple resources are provided 
through multiple layers of networks~\cite{Kullman1979,Rinaldi2001,Buldyrev2010,Min2014resource}. 
For example, computer-controlled systems function properly only when the Internet and 
power grids are supplied interdependently~\cite{Rinaldi2001,Buldyrev2010}
In addition, for a city to function properly, resources such as water, gas, and electricity 
must be supplied by  separate channels~\cite{Kullman1979,Min2014resource,Min2015}. 
For this reason, the mutual connectivity of multiple layers of networks has received much 
attention for several years~\cite{Buldyrev2010,Son2012,Baxter2012,Min2014resource,Kivela2014,Lee2015,Lee2018}.
A mutually-connected component, which is a central concept in mutual percolation, 
is defined as a set of nodes that every node pair in the component 
has at least one path, composed of nodes within the same cluster, in each and every 
layer of networks~\cite{Buldyrev2010,Baxter2012}.
Pioneering works on the mutual percolation have studied the robustness of interdependent networks by
using the notion of a mutually connected giant component and showed that an abrupt transition 
can appear between percolating and non-percolating
phases~\cite{Buldyrev2010,Baxter2012,Son2012,Min2014Network}. 
These studies imply that a small perturbation can cause an abrupt collapse of the entire 
system, posing a potentially catastrophe to complex systems~\cite{Buldyrev2010,Son2012,Min2014resource,Lee2016,Zhou2014,Baxter2014,Hackett2016}

Since many real-world systems are embedded in low dimensions, percolation problems on 
multi-layered two-dimensional lattices have been also of interest~\cite{Son2011,Li2012,Bashan2013}. 
Unlike multiplex networks, the 
mutually connected giant cluster on two-dimensional lattices emerges continuously 
as the probability of bond occupation increases~\cite{Son2011,Li2012}. However, the 
studies on the critical behaviors of the mutual percolation on two-dimensional lattices 
yielded conflicting results~\cite{Son2011,Berezin2013,Son2013,Grassberger2015}.
Son {\it et al.} demonstrated that the percolation transition for interdependent diluted 
lattices belongs to a different universality class with a larger order parameter exponent 
than that of ordinary percolation~\cite{Son2011,Son2013,Grassberger2015}.
However, another study reported that the mutual percolation on two-dimensional lattices belongs 
to the same universality class of ordinary percolation~\cite{Berezin2013}.
In addition to mutual percolation, several percolation problems in multi-layer lattices
have been also studied such as history dependent percolation~\cite{Jovanovic1994,Hu2020,Li2020},
and percolation in porous media~\cite{Cai1990,Hansen1993}.

From a perspective different from topological aspect, there have been 
studies on the viability of nodes as a consequence of cascading dynamics following
the iterative activations or deactivations of nodes \cite{Min2014resource,Min2015,Baxter2014}. 
A model for the viability deals with systems that demand more than one type of vital resource 
to be produced and distributed by source nodes in multiplex networks \cite{Min2014resource}. 
Viable nodes in this model are identified by using two different dynamical processes: 
cascade of activations ($\mathrm{CA}$) and deactivations ($\mathrm{CD}$).
The $\mathrm{CA}$ process starts with all the sites in the deactivated state, and finds mutually-connected clusters 
through the diffusion of activations from the source nodes; in contrast, the $\mathrm{CD}$ process finds 
mutually-connected clusters by iteratively removing any sites that are not mutually-connected. 
On multiplex networks, these processes can produce two different final configurations of
viable nodes, corresponding to different stable solutions of a single mean-field 
equation \cite{Min2014resource,Baxter2014}.
Viability is strongly related to mutual percolation on multiplex networks 
because the dynamical consequence of $\mathrm{CD}$ becomes identical to mutual percolation 
in the limit when the fraction of initial source nodes goes to be zero. 
While cascading dynamics on multiplex networks has been studied \cite{Min2014resource,Min2015},
it has not been explored on low-dimensional systems, especially
the critical behaviors of cascading dynamics.

In this work, we study the critical phenomena of cascading dynamics on 
multiplex two-dimensional lattices using extensive Monte-Carlo simulations. We found that 
two cascading processes, $\mathrm{CA}$ and $\mathrm{CD}$, can lead to different 
critical behaviors in viable clusters statistics.
While $\mathrm{CD}$ exhibits the same set of critical exponents reported for mutual 
percolation on interdependent networks \cite{Son2011}, $\mathrm{CA}$ exhibits the critical 
behaviors akin to the ordinary percolation on two-dimensional lattices. 
In addition, we computed the susceptibility of cascading dynamics by using the 
ghost field concept~\cite{Kasteleyn1969,Reynolds1977} and confirmed that the hyperscaling 
relation is satisfied.

\section{Model}

\begin{figure}
\includegraphics[width=1\linewidth]{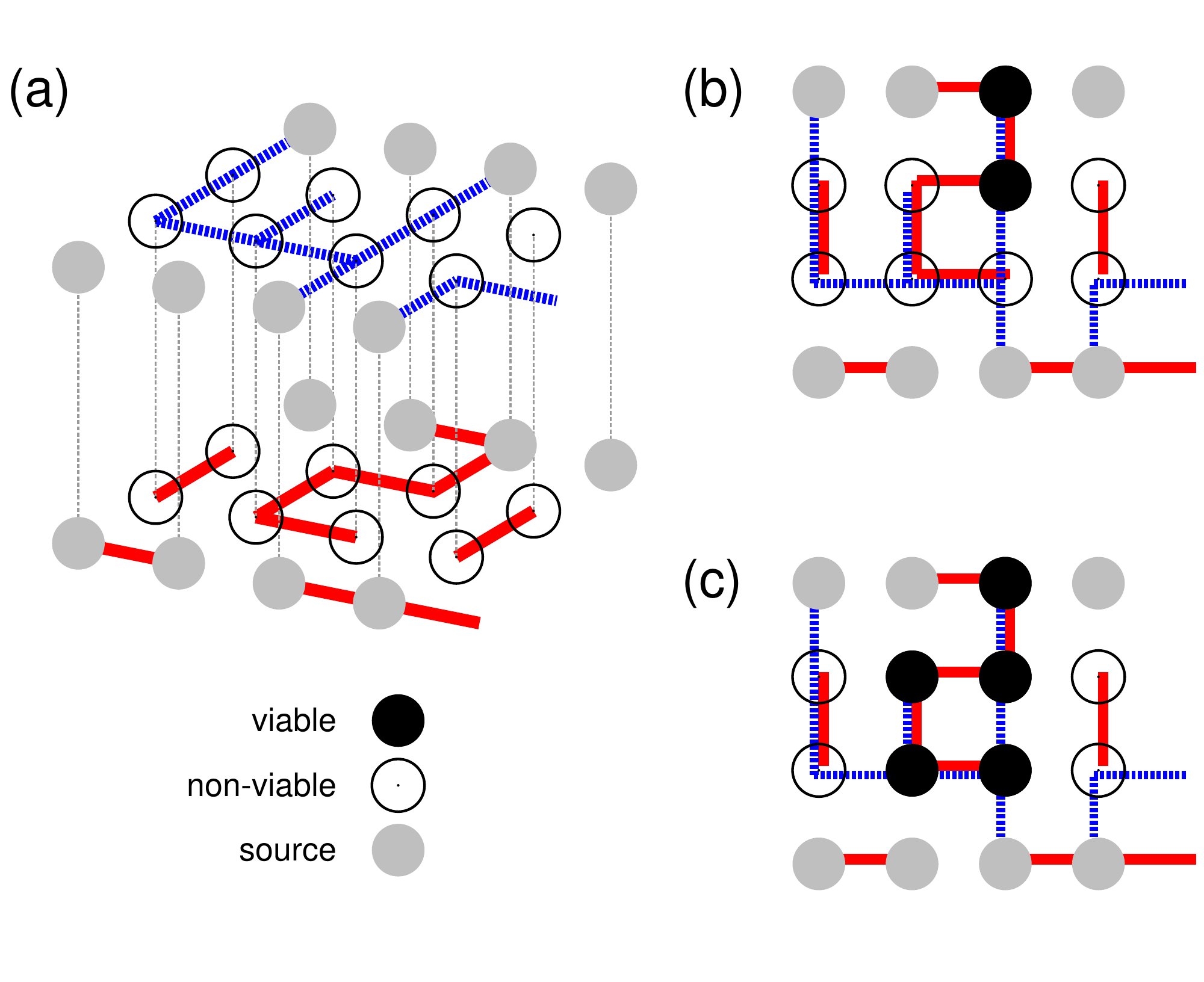}
\caption{
(a) A multiplex lattice consists of double-layered two-dimensional diluted lattices.
(b,c) Examples of the final configurations of (b) $\mathrm{CA}$ and (c) $\mathrm{CD}$ processes are 
depicted. Viable, non-viable, and source sites are respectively denoted by black filled 
circles, black open circles, and grey filled circles.
}
\end{figure}

Let us imagine an example for cascading processes on a multiplex lattice.
A site needs two resources, say $\mathcal{A}$ and $\mathcal{B}$, to be viable, e.g., a city 
requires water and electricity for functioning. Each resource is delivered along the 
corresponding layer of the lattice. Initially, the resources are generated from a few source sites 
in each layer. The important point is that only a viable site can convey resources to the neighboring 
sites along connected bonds. In this setting, we consider the following two processes for finding 
viable sites from the given lattice structures and initial source configurations. 
i) Assume that all sites except the source sites are initially in the non-viable state. Sites that 
satisfy the viable condition change into the viable state. These viable sites are able to convey 
resources to their neighbors. There will be sites that change to the viable state in turn
and we call this process the cascade of activation ($\mathrm{CA}$). 
ii) On the other hand, one may start at the beginning when all sites are putatively in a viable state.
In this case, sites that do not satisfy the viable condition must change from the viable to non-viable state
iteratively until no further deactivation is necessary.
This process is called the cascade of deactivation ($\mathrm{CD}$).

The specific simulation rules are as follows. There are two layers of two-dimensional 
lattices, say $\mathcal{L}_A$ and $\mathcal{L}_B$ with  $N=L \times L$,
where $L$ denotes the size	of the lattice.
Each bond in the lattice is occupied with probability $p$, and 
empty with probability $1-p$. On top of multiplex diluted lattices,
resources $\mathcal{A}$ and $\mathcal{B}$ are distributed at their source sites. 
In this study, we assume that the sources of resources $\mathcal{A}$ and $\mathcal{B}$ are located at 
every site on the boundary of lattices $\mathcal{L}_A$ and $\mathcal{L}_B$, respectively [Fig.~1(a)]. 
We use cylindrical boundary conditions. 
Then, each site on a multiplex lattice is viable only if both resources $\mathcal{A}$ and $\mathcal{B}$ 
are supplied. Moreover, only viable sites can deliver resources to their neighbors connected by occupied bonds.
Otherwise a site becomes non-viable and cannot convey resources further. 
Note that resources $\mathcal{A}$ and $\mathcal{B}$ are respectively supplied only through 
the chain of viable sites on $\mathcal{L}_A$ and $\mathcal{L}_B$. 
Using the two processes, $\mathrm{CA}$ and $\mathrm{CD}$, we find  viable sites 
in the steady state. We then identify the viable clusters as a set of viable sites 
for every pair of sites in the cluster has at least one path composed of viable 
sites within each and every layer of the lattice. 
We also define the fraction of the giant viable cluster as $V_{\mathrm{CA}}$ for 
$\mathrm{CA}$ as shown in Fig.~1(b), and $V_{\mathrm{CD}}$ for 
$\mathrm{CD}$ as shown in Fig.~1(c).

\section{Results}

We conducted Monte Carlo simulations for $\mathrm{CA}$ and 
$\mathrm{CD}$ processes on multiplex lattices with various sizes $L$.
We then identify the configuration of the viable clusters and analyze
their critical behaviors at the steady state. 
The fractions $V$ of the giant viable cluster for the two processes are 
in general different from each other as shown in Fig.~2(a), as 
in the case of multiplex networks \cite{Min2014resource}. 
The giant viable cluster for $\mathrm{CD}$ process appears at a smaller $p$
compared with that for $\mathrm{CA}$. In addition, the fraction $V_{\mathrm{CD}}$ of 
the giant viable cluster for $\mathrm{CD}$ is generally larger than $V_{\mathrm{CA}}$.

The critical behaviors of percolation problems are typically characterized by the divergence
of the average cluster size $\chi$ and correlation length $\xi$ at the critical point $p_c$
in the thermodynamic limit $L\rightarrow \infty$. Near the critical point, the order parameter $V$,
susceptibility $\chi$, and correlation length $\xi$ follow power laws with the 
critical exponents $\beta$, $\gamma$, and $\nu$ as follows:
\begin{align}
V(p) \sim (p-p_c)^\beta, \\
\chi(p) \sim |p-p_c|^{-\gamma}, \\
\xi(p) \sim |p-p_c|^{-\nu}. 
\end{align}
In order to obtain the critical exponents, we use the conventional 
scaling ansatz for finite-size systems \cite{Stauffer2018},
\begin{align}
V(p,L)=L^{-\beta/\nu}f_V[(p-p_c)L^{1/\nu}],
\label{fsso}
\\
\chi(p,L) = L^{\gamma / \nu}f_{\chi} [(p-p_c(L))L^{1/\nu}].
\label{fsss}
\end{align}
where $p_c(L)$ is the location of the peak of susceptibility.

\begin{figure}
\includegraphics[width=\linewidth]{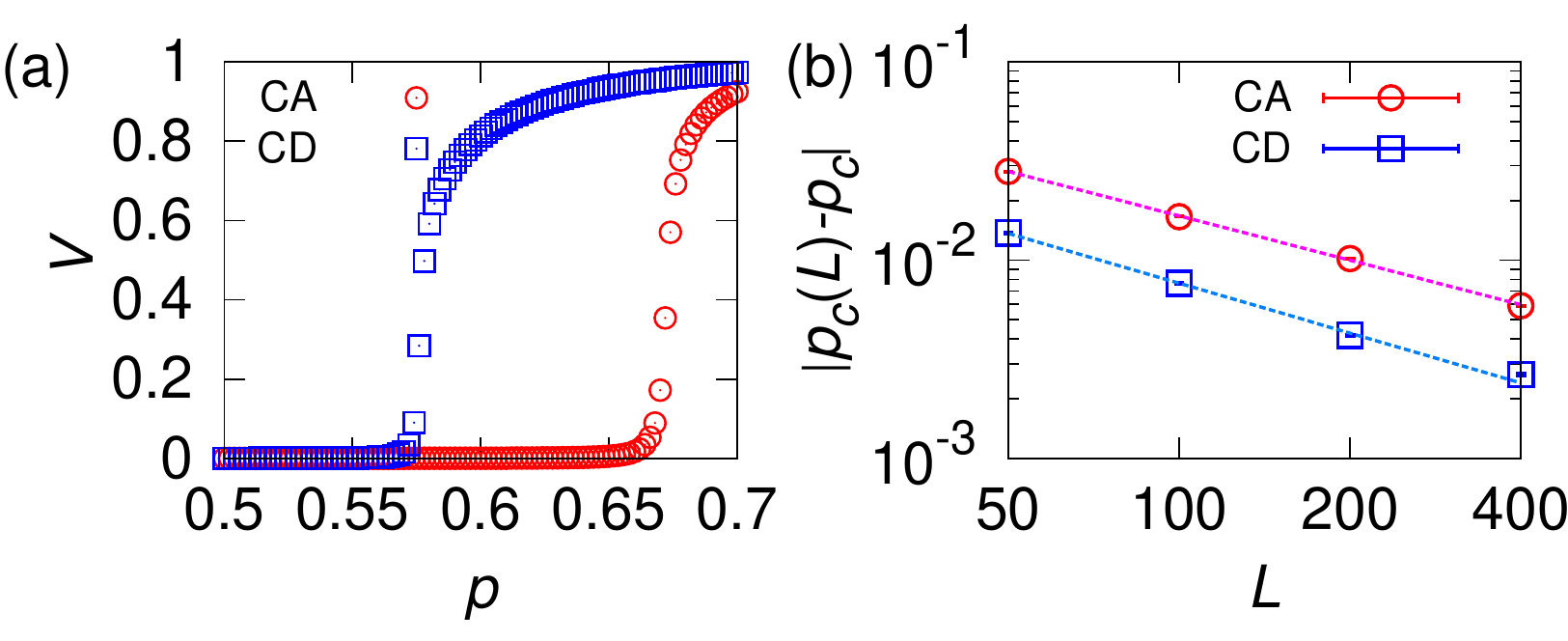}
\caption{
(a) Numerical results of the fraction $V$ for the $\mathrm{CA}$ (red) 
and $\mathrm{CD}$ (blue) processes as a function of the bond occupation probability $p$ 
on multiplex lattices with $L=800$, averaged over $100$ different realizations are shown. 
(b) The relation $|p_c-p_c(L)| \sim L^{-1/\nu}$ for $\mathrm{CA}$ (red) and $\mathrm{CD}$ (blue) is shown.
The points represent numerical simulation results averaged over $400$ samples. 
}
\label{fig2}
\end{figure}

While computing the fraction $V$ of the giant  viable cluster is straightforward,
how to measure susceptibility $\chi$ is far from trivial.
For the ordinary percolation, the susceptibility $\chi$ is directly measured by
the average size of the finite clusters.
In our model, however, the quantity corresponding to the susceptibility is not 
readily provided by the size of the average finite viable clusters.
To obtain the susceptibility in $\mathrm{CA}$ and $\mathrm{CD}$, we use 
the concept of ``ghost field'' analogous to  magnetic 
susceptibility \cite{Kasteleyn1969,Reynolds1977,Fortuin1972}.
We add one additional ghost site that is connected to each site in the system with
a probability $H$. By applying a small amount of $H$ 
corresponding to the external magnetic field in spin systems, we measure
the change in order parameter $V$ such that $\chi = \partial V/ \partial H$.
In practice, we choose one site that does not belong to the giant viable cluster. 
Next, we assume that this site has become part of the giant viable cluster 
through the connection to the ghost site. We then identify a set of sites that 
newly entered the giant viable cluster because of the presence of the ghost site.
We call the set of sites as ``susceptible cluster'' and measure the size of the 
susceptible cluster as susceptibility. 
In the numerical simulations, we attempted to select every site that does not belong 
to the giant viable cluster and measured the distribution $\phi(s)$ that a site belongs to
a susceptible cluster with size $s$. The susceptibility $\chi$ is obtained as
\begin{equation}
\chi(p,L)= \frac{\sum_{s}  s\phi(s,p)}{\sum_{s} \phi(s,p)},
\label{eq:sus}
\end{equation}
where the sum is over all cluster sizes excluding the giant viable cluster.

We estimated the viability threshold and critical exponent $\nu$ 
for both $\mathrm{CA}$ and $\mathrm{CD}$ using a finite size scaling ansatz.
We first obtained the  threshold as $p_c^\mathrm{CA} \simeq 0.6743$ 
and $p_c^\mathrm{CD}\simeq 0.5761$.
As in the case of $\mathrm{CA}$ and $\mathrm{CD}$ in multiplex networks \cite{Min2014resource}, 
the giant viable clusters in $\mathrm{CA}$ processes emerges at a larger $p$ compared to $\mathrm{CD}$. 
Next, we obtained the exponent of the correlation length $\nu$ from the relation
$|p_c(L) - p_c| \sim L^{-1/\nu}$ as shown in Fig.~2(b).  
Here, $p_c(L)$ is a threshold with size $L$ and is estimated by the value 
of $p$ at the maximum value of susceptibility $\chi$.
The fitted values of $\nu$ for $\mathrm{CA}$ and $\mathrm{CD}$ are respectively $1.342(3)$ and $1.197(7)$.

\begin{figure}[t]
\includegraphics[width=\linewidth]{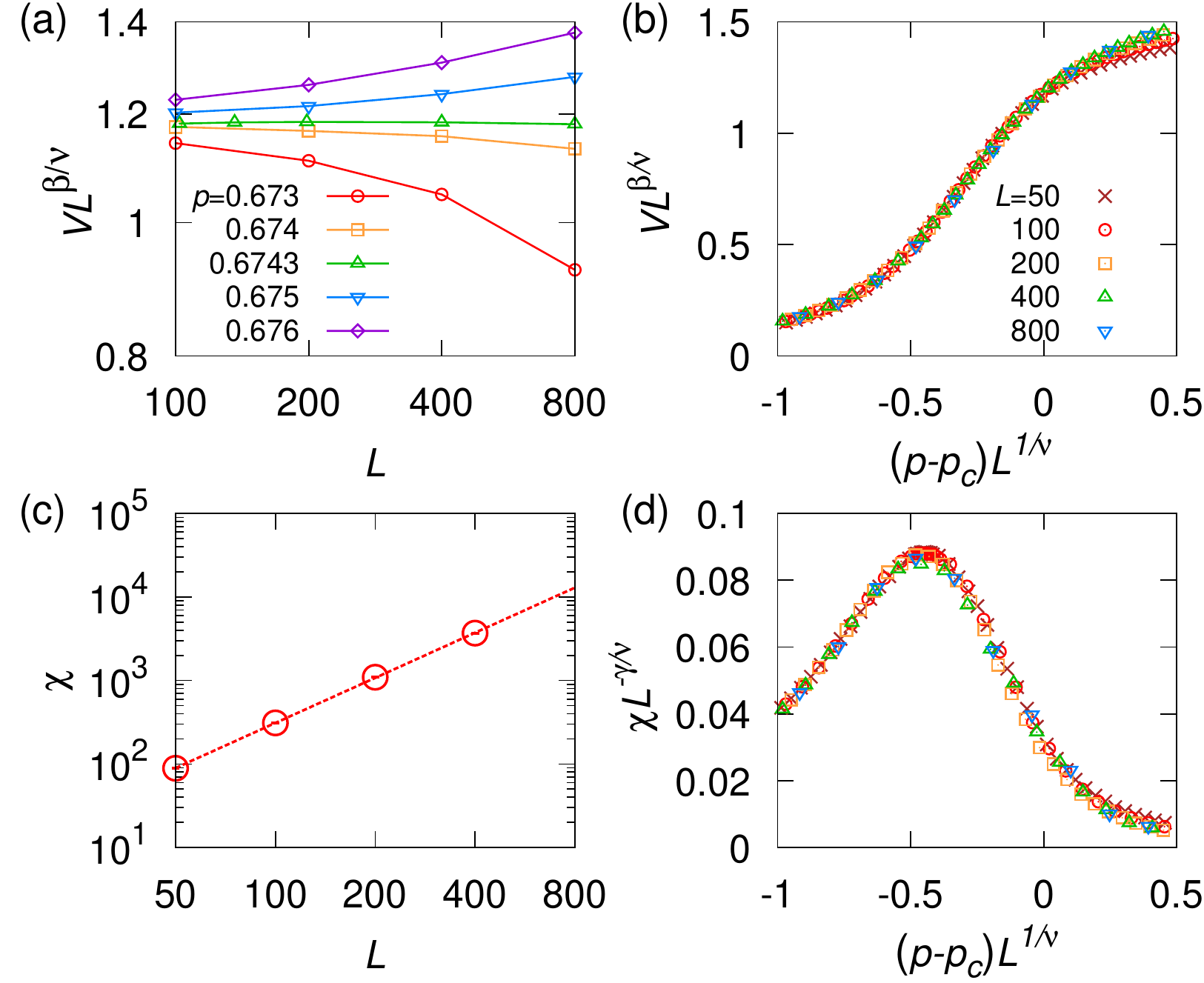}
\caption{
(a) Log-log plot of $V_\mathrm{CA} L^{\beta/\nu}$ at $p_c^\mathrm{CA}$ with respect to $L$ for various $p$.
(b) Data collapse of the scaled order parameter $V_\mathrm{CA} L^{\beta/\nu}$ vs. 
$(p-p_c)L^{1/\nu}$ with $\beta=0.137(1)$ and $\nu=1.342(3)$.
(c) Log-log plot of $\chi_\mathrm{CA}$ at $p_c(L)$ with respect to $L$, showing the 
relation $\chi \sim L^{\gamma/\nu}$.
(d) Data collapse of the scaled susceptibility $\chi_\mathrm{CA} L^{-\gamma/\nu}$ vs.
$(p-p_c)L^{1/\nu}$ with $\gamma=2.41(1)$ and $\nu=1.342(3)$.
Numerical results of $\mathrm{CA}$ were obtained for various lattice sizes $L$ from 50 to 800, averaged over 
400 realizations. 
}
\label{fig3}
\end{figure}

\begin{figure}[t]
\includegraphics[width=\linewidth]{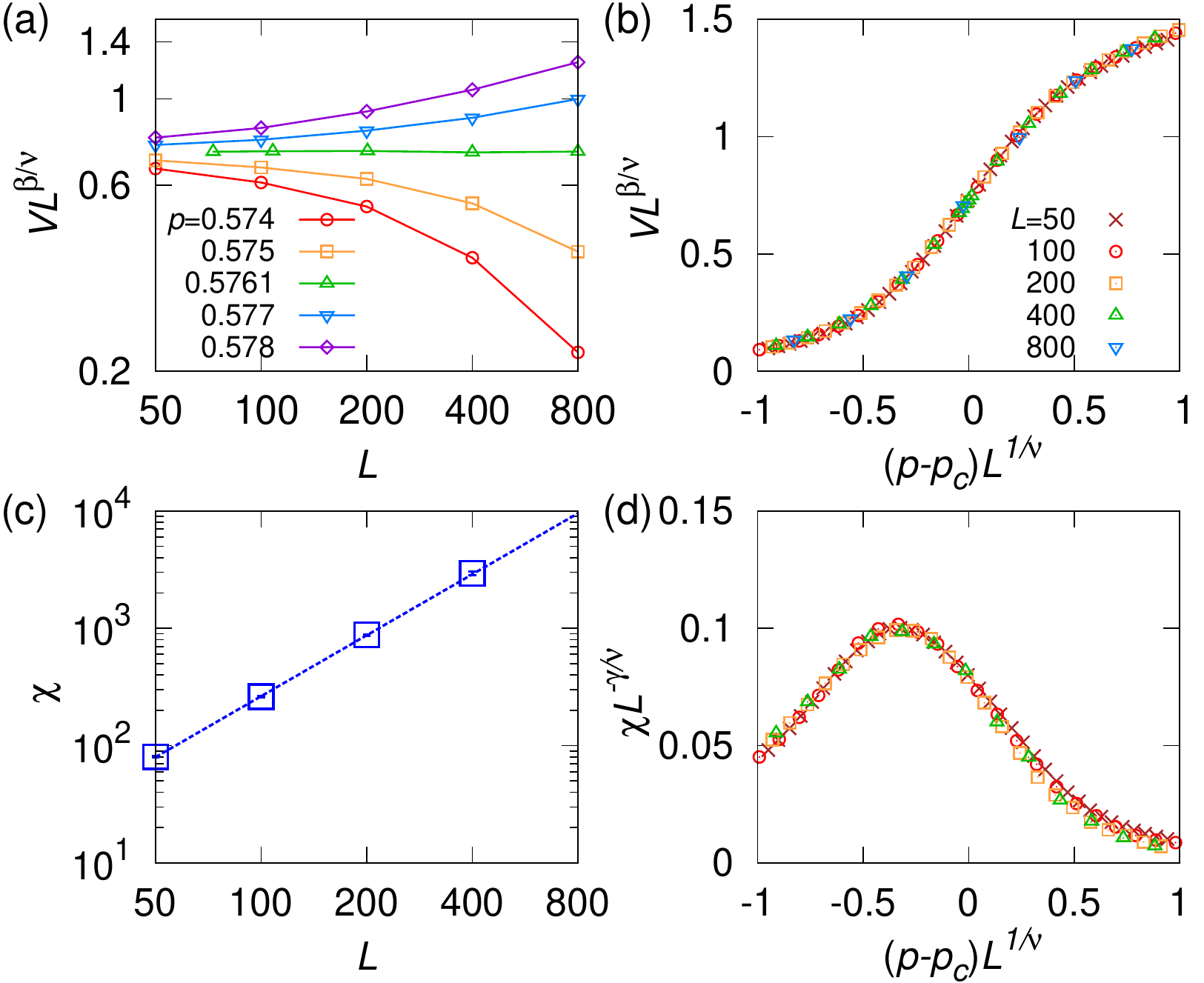}
\caption{
(a) Log-log plot of $V_\mathrm{CD} L^{\beta/\nu}$ at $p_c^\mathrm{CD}$ with respect to $L$ for various $p$.
(b) Data collapse of the scaled order parameter $V_\mathrm{CD} L^{\beta/\nu}$ vs. 
$(p-p_c)L^{1/\nu}$ with $\beta=0.163(2)$ and $\nu=1.197(7)$.
(c) Log-log plot of $\chi_\mathrm{CD}$ at $p_c(L)$ with respect to $L$, showing the 
relation $\chi \sim L^{\gamma/\nu}$.
(d) Data collapse of the scaled susceptibility $\chi_\mathrm{CD} L^{-\gamma/\nu}$ vs.
$(p-p_c)L^{1/\nu}$ with $\gamma=2.07(1)$ and $\nu=1.197(7)$.
Numerical results of $\mathrm{CD}$ were obtained for various lattice sizes $L$ from 50 to 800, averaged over 
400 realizations. 
}
\label{fig4}
\end{figure}

We then obtained $\beta/\nu$ and $\gamma/\nu$ using the scalings respectively
$V \sim L^{-\beta/\nu}$ and $\chi \sim L^{\gamma/\nu}$. 
The results of critical behaviors for $\mathrm{CA}$ processes 
are shown in Fig.~3.
Figure 3(a) shows the scaling of $V \sim L^{-\beta/\nu}$ with various values of $p$.
The scaling shows the estimates of the exponents $\beta=0.137(1)$ and $\nu=1.342(3)$. 
Data collapse curves of $V L^{\beta/\nu}$ with respect to $(p-p_c)L^{1/\nu}$
in Fig.~3(b) confirm the estimated values of the critical exponents.
Figure 3(c) shows the relationship $\chi \sim L^{\gamma/\nu}$ with $\gamma/\nu = 1.794(6)$
which is confirmed by the finite-size-scaled data collapse of $\chi L^{-\gamma/ \nu}$
with respect to $(p-p_c)L^{1/\nu}$, as shown in Fig.~3(d).
We found that the critical exponents for $\mathrm{CA}$ processes are consistent  
within error bars for ordinary bond percolation on two-dimensional lattices.

For $\mathrm{CD}$ processes, we also obtain $\beta$ and $\gamma$ using scaling
$V \sim L^{-\beta/\nu}$ and $\chi \sim L^{\gamma/\nu}$ as shown in Fig.~4.
The estimated values of the critical exponents for the $\mathrm{CD}$ processes are 
$\beta=0.163(2)$ and $\nu=1.197(7)$ [Fig.~4(a)]. 
In addition, figure 4(c) exhibits the relationship $\chi \sim L^{\gamma/\nu}$ with $\gamma/\nu = 1.73(1)$.
We also confirm the critical exponents
by using the data collapse of $V L^{\beta/\nu}$ with respect to $(p-p_c)L^{1/\nu}$ [Fig.~4(b)]
and $\chi L^{-\gamma/ \nu}$ with respect to $(p-p_c)L^{1/\nu}$ [Fig.~4(d)].
The values of the exponents $\beta$, $\gamma$, and $\nu$ are distinct from 
those for $\mathrm{CA}$. 
In addition, the critical behaviors of $\mathrm{CD}$ are 
consistent within error bars of the mutual percolation on two-dimensional lattices reported 
in \cite{Son2011,Grassberger2015}.

\begin{figure}[t]
\includegraphics[width=\linewidth]{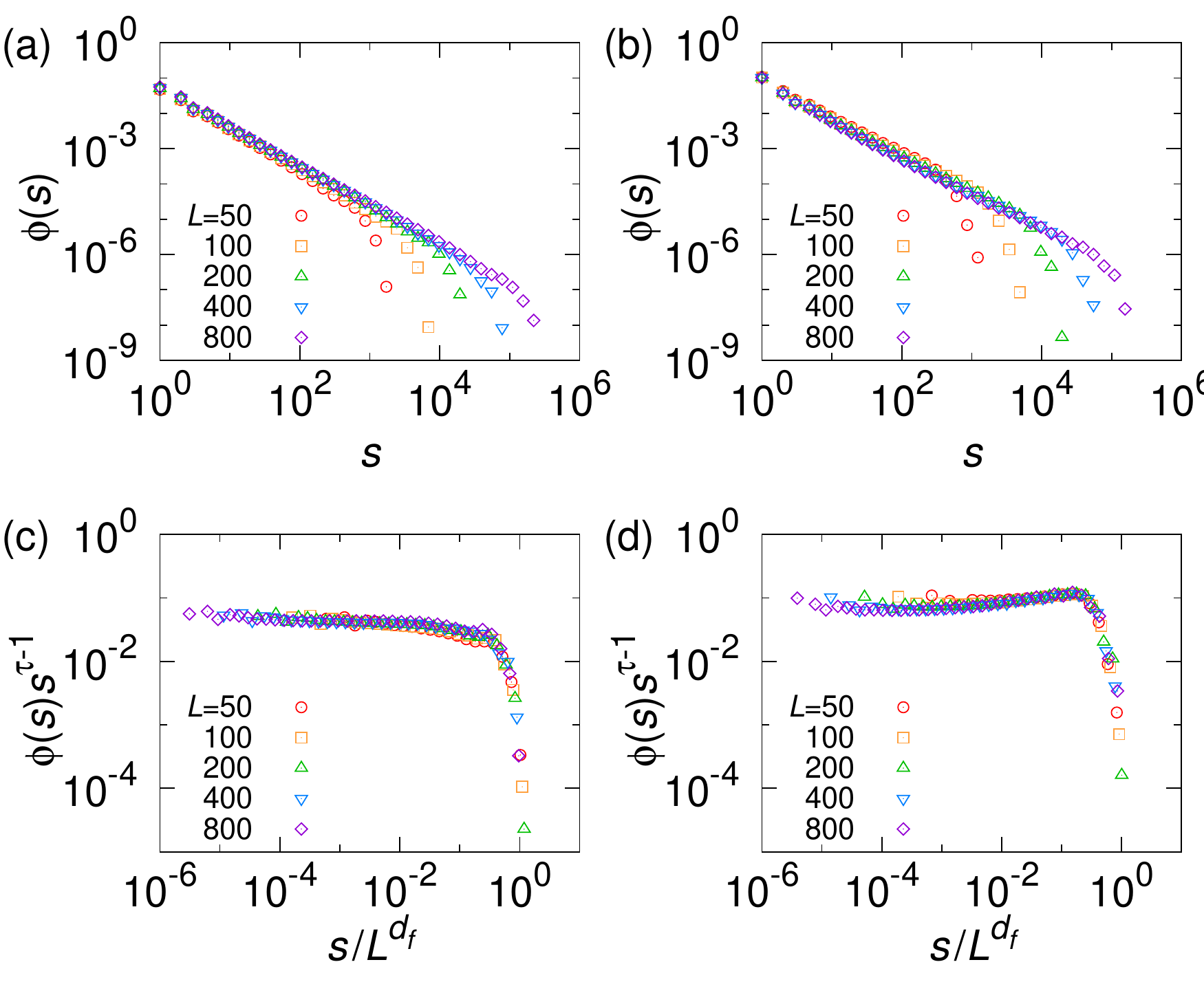}
\caption{Probability distributions $\phi(s)$ that a node belongs to a susceptible cluster
with size $s$ for (a) $\mathrm{CA}$ and (b) $\mathrm{CD}$.
Data collapses of $\phi(s) s^{\tau-1}$ vs. $s/L^{d_f}$ 
for (c) $\mathrm{CA}$ with $\tau=2.06(3)$ and $d_f=1.8976$ and 
for (d) $\mathrm{CD}$ with $\tau=2.07(1)$ and $d_f=1.864$.
The fractal dimension $d_f$ is estimated by $d_f = d - \beta/\nu$.
Monte-carlo simulations were performed at the critical point $p=0.6743$ for 
$\mathrm{CA}$ and $p=0.5761$ for $\mathrm{CD}$.
Numerical results were obtained for various lattice sizes $L$ from 50 to 800, averaged over 
400 realizations. 
}
\label{fig6}
\end{figure}

We also obtain the size distribution $\phi(s)$ that a site belongs to 
susceptible clusters with size $s$. At $p_c$, the size distribution 
follows the power-law form $\phi(s) \sim s^{1-\tau}$ in the thermodynamic limit.
We apply a scaling ansatz to the distribution 
\begin{align}
\phi(s) = s^{1-\tau} g(s/L^{d_f}) 
\label{fsscls} 
\end{align}
where $g$ is a scaling function and $d_f$ is the fractal dimension
given by $d_f = d - \beta/\nu$ where $d$ is the spatial dimension of lattices.
Figures~\ref{fig6}(a,b) show the power-law decay of $\phi(s)$ at $p_c$ 
for (a) $\mathrm{CA}$ and (b) $\mathrm{CD}$, respectively. We found the estimate of 
critical exponents as $\tau=2.06(3)$ for $\mathrm{CA}$ and $\tau=2.07(1)$ for $\mathrm{CD}$.
We confirm the scaling ansatz using data collapses, as shown in 
Figs.~\ref{fig6}(c,d).

The obtained values of the threshold and critical exponents are presented in Table 1. 
We observed that the critical exponents for $\mathrm{CA}$ and $\mathrm{CD}$ 
are distinct.  Moreover, all critical exponents of $\mathrm{CA}$ 
are consistent within error bars with ordinary bond percolation on two-dimensional lattices.
However, the critical exponents of $\mathrm{CD}$ are consistent with those of the mutual 
percolation on two-dimensional lattices, which shows a discrepancy with $\mathrm{CA}$.
In addition, the sets of exponents for both $\mathrm{CA}$ and $\mathrm{CD}$ 
satisfy the hyperscaling relations $d\nu=2\beta+\gamma$ and $\tau=d/d_f +1$.

\begin{table}[b]
\caption{Critical points and exponents}
\begin{ruledtabular}
\begin{tabular}{c|cc|cc}
\multirow{3}{*}~&\multirow{3}{*}{$\mathrm{CA}$} &Ordinary&\multirow{2}{*} {$\mathrm{CD}$}&Mutual\\
 ~& &Percolation& &Percolation \cite{Grassberger2015}\\
\hline
$p_c$&0.6743(1)& 1/2 & 0.5761(1)&0.576132(5)\\
$\nu$& 1.342(3)& 4/3 & 1.197(7)& 1.200(7)\\
$\beta$& 0.137(1)& 5/36 & 0.163(2)& 0.163(2)\\
$\gamma$& 2.41(1)& 43/18 & 2.07(2)& $\cdot$ \\
$\tau$&  2.06(3) & 187/91 & 2.07(1) & $\cdot$ \\
\hline
$1/\nu$& 0.745(2)& 3/4 & 0.836(5)& 0.833 \\
$\beta/\nu$& 0.1024(5)& 5/48 & 0.136(1)& 0.1358 \\
$\gamma/\nu$& 1.794(6)& 43/24 & 1.73(1)& 1.728 \\
\end{tabular}
\end{ruledtabular}
\end{table}

\begin{figure}[t]
\includegraphics[width=\linewidth]{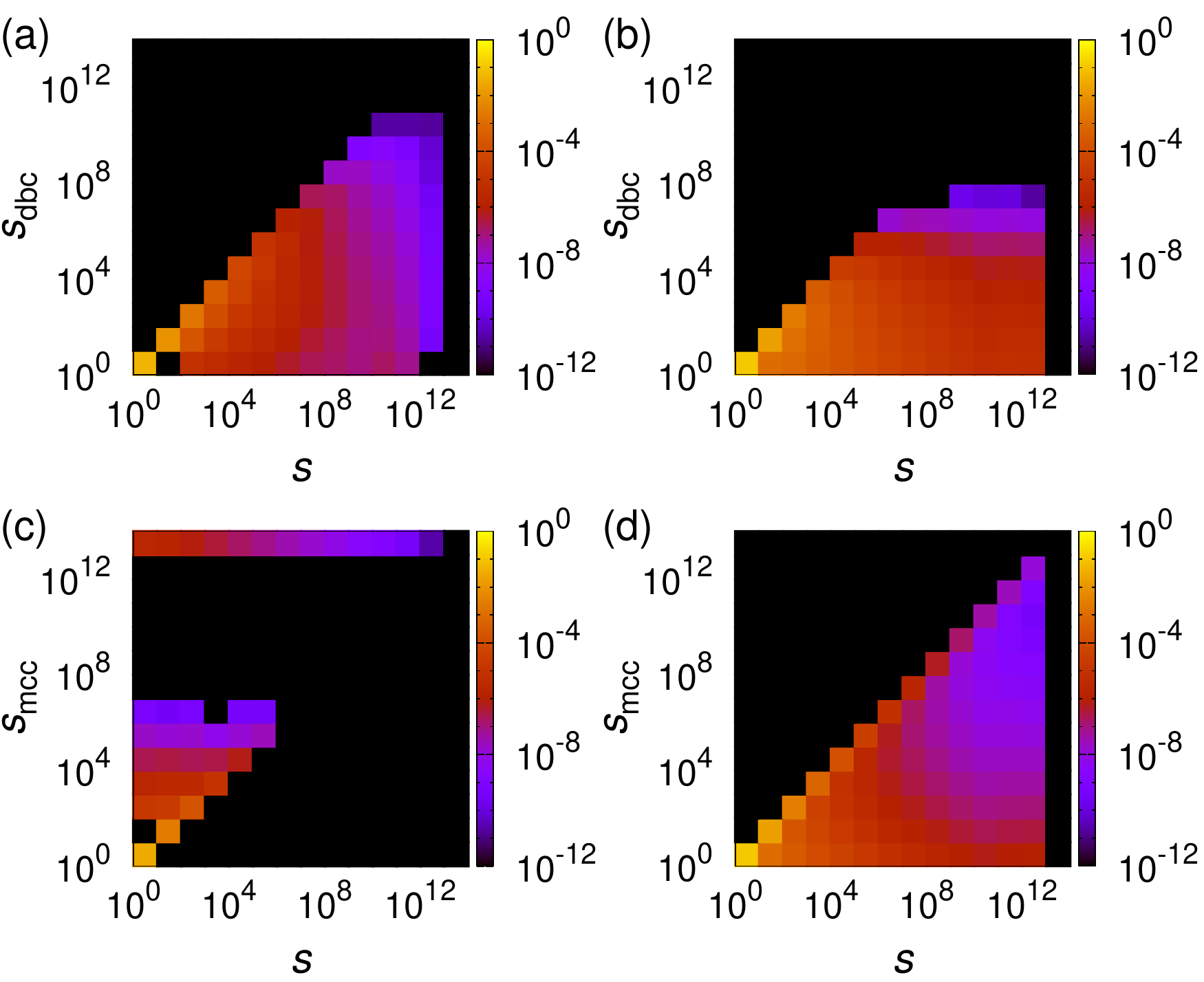}
\caption{Joint probability distributions $J(s,s_\textrm{dbc})$ where $s$ is the size of susceptible clusters
and $s_\textrm{dbc}$ is that of double-bond clusters are shown for (a) $\mathrm{CA}$ and (b) $\mathrm{CD}$.
Joint probability distributions $J(s,s_\textrm{mcc})$ where $s_\textrm{mcc}$ is the size of mutually 
connected clusters are shown for (c) $\mathrm{CA}$ and (d) $\mathrm{CD}$.
Numerical results are calculated at the critical points $p=0.6743$ for $\mathrm{CA}$ and $p=0.5761$ 
for $\mathrm{CD}$ with $L=100$, averaged over $5000$ realizations.
}
\label{fig5}
\end{figure}

To understand the differences between the underlying mechanisms forming susceptible 
clusters, we examine the joint probability distribution between the sizes of susceptible clusters $s$ and 
that of double-bond clusters $s_\textrm{dbc}$, and that of mutually connected clusters $s_\textrm{mcc}$. 
A double bond refers to a connection in which a bond between two adjacent sites exists in both layers,
and the double-bond cluster stands for a set of sites that are connected solely via double bonds.
The mutually connected component is defined as a set of sites that every pair of sites in the cluster
has at least one path within each and every layer of the lattice.

Figures~\ref{fig5}(a,b) show the joint probability distribution $J(s,s_\textrm{dbc})$ for (a) $\mathrm{CA}$
and (b) $\mathrm{CD}$ to examine the effect of double bonds in forming susceptible clusters.
The joint probability distributions are computed at the critical point, $p=0.6743$ for $\mathrm{CA}$
and $p=0.5761$ for $\mathrm{CD}$.
We found that the size of the double-bond clusters are strongly correlated 
to that of the susceptible clusters for $\mathrm{CA}$. This means that the double bonds play a 
dominant role in the $\mathrm{CA}$ process. For $\mathrm{CD}$ process, however, small
susceptible clusters are formed by double bonds, while large susceptible clusters are not.
We found that large-sized susceptible clusters in $\mathrm{CD}$ are constructed by 
mutually connected clusters as shown in Fig.~\ref{fig5}(d). 
On the other hand, for the $\mathrm{CA}$ process there is already a mutually connected 
giant component near the critical point as shown in Fig.~\ref{fig5}(c). 
Therefore, mutually connected clusters cannot contribute crucially the emergence of 
the large-sized susceptible clusters in $\mathrm{CA}$ process. 
In conclusion, we found that large-sized susceptible clusters are mainly constructed by double
bonds in the $\mathrm{CA}$ process whereas they are formed by mutual connections 
in the $\mathrm{CD}$ process.

\section{Discussion}

In this work, we study the critical phenomena of cascading dynamics on multiplex lattices
considering two different processes to identify viable clusters.
We found that viable clusters identified by the $\mathrm{CA}$ and $\mathrm{CD}$ processes 
exhibits different critical behaviors. The $\mathrm{CA}$ process exhibits 
critical phenomena consistent to ordinary bond percolation but the $\mathrm{CD}$ 
process exhibits those to mutual percolation on multiplex lattices. 
In addition, we explicitly calculate susceptibility by using the ghost field.
Our study shows that viability that requires multiple connectivity can be maintained 
by diverse mechanisms in a multiplex low-dimensional system.
The cascading dynamics that we consider here is an example of cooperative couplings
between multiple layers. Similar cooperative interactions can be realized in real-world 
systems in a variety of ways such as cooperative infections \cite{Chen2013,Min2020}, social 
contagions \cite{Lee2014,Min2018}, and interdependencies in multi-layered 
systems \cite{Buldyrev2010,Pocock2012}. In this regard, our research shows 
a glimpse of the complexity generated by cooperative interactions 
in multiplex low-dimensional systems. Further studies may be required to examine 
the cascading dynamics on an arbitrary number of layers and
partially coupled layers.

\begin{acknowledgments}
This work was supported by the National Research Foundation of Korea (NRF) grants funded by the Korea 
government (MSIT) (No.~2020R1I1A3068803 (BM) and No.~2020R1A2C2003669 (K-IG)). 
\end{acknowledgments}

\bibliography{./cacd}

\begin{thebibliography}{10}
\expandafter\ifx\csname url\endcsname\relax
  \def\url#1{\texttt{#1}}\fi
\expandafter\ifx\csname urlprefix\endcsname\relax\def\urlprefix{URL }\fi
\expandafter\ifx\csname href\endcsname\relax
  \def\href#1#2{#2} \def\path#1{#1}\fi

\bibitem{Kullman1979}
D.~E. Kullman, The utilities problem, Mathematics Magazine 52~(5) (1979)
  299--302.

\bibitem{Rinaldi2001}
S.~Rinaldi, J.~Peerenboom, T.~Kelly, Identifying, understanding, and analyzing
  critical infrastructure interdependencies, IEEE Control Systems Magazine
  21~(6) (2001) 11--25.
\newblock \href {https://doi.org/10.1109/37.969131}
  {\path{doi:10.1109/37.969131}}.

\bibitem{Buldyrev2010}
S.~V. Buldyrev, R.~Parshani, G.~Paul, H.~E. Stanley, S.~Havlin, {Catastrophic
  cascade of failures in interdependent networks}, Nature 464~(7291) (2010)
  1025--1028.
\newblock \href {https://doi.org/10.1038/nature08932}
  {\path{doi:10.1038/nature08932}}.

\bibitem{Min2014resource}
B.~Min, K.-I. Goh, Multiple resource demands and viability in multiplex
  networks, Phys. Rev. E 89~(4) (2014) 040802(R).

\bibitem{Min2015}
B.~Min, S.~Lee, K.-M. Lee, K.-I. Goh, Link overlap, viability, and mutual
  percolation in multiplex networks, Chaos, Solitons, and Fractals 72 (2015)
  49.

\bibitem{Son2012}
S.-W. Son, G.~Bizhani, C.~Christensen, P.~Grassberger, M.~Paczuski, Percolation
  theory on interdependent networks based on epidemic spreading, Europhysics
  Letters 97~(1) (2012) 16006.
\newblock \href {https://doi.org/10.1209/0295-5075/97/16006}
  {\path{doi:10.1209/0295-5075/97/16006}}.

\bibitem{Baxter2012}
G.~J. Baxter, S.~N. Dorogovtsev, A.~V. Goltsev, J.~F.~F. Mendes, {Avalanche
  Collapse of Interdependent Networks}, Phys. Rev. Lett. 109~(24) (2012)
  248701.
\newblock \href {https://doi.org/10.1103/PhysRevLett.109.248701}
  {\path{doi:10.1103/PhysRevLett.109.248701}}.

\bibitem{Kivela2014}
M.~Kivela, A.~Arenas, M.~Barthelemy, J.~P. Gleeson, Y.~Moreno, M.~A. Porter,
  Multilayer networks, Journal of Complex Networks 2~(3) (2014) 203--271.
\newblock \href {https://doi.org/10.1093/comnet/cnu016}
  {\path{doi:10.1093/comnet/cnu016}}.

\bibitem{Lee2015}
K.-M. Lee, B.~Min, K.-I. Goh, Towards real-world complexity: an introduction to
  multiplex networks, The European Physical Journal B 88~(48) (2015) 48.

\bibitem{Lee2018}
D.~Lee, B.~Kahng, Y.~S. Cho, K.~I. Goh, D.~S. Lee, Recent advances of
  percolation theory in complex networks, Journal of the Korean Physical
  Society 73~(2) (2018) 152--164.

\bibitem{Min2014Network}
B.~Min, S.~D. Yi, K.-M. Lee, K.-I. Goh, {Network robustness of multiplex
  networks with interlayer degree correlations}, Phys. Rev. E 89~(4) (2014)
  042811.
\newblock \href {https://doi.org/10.1103/PhysRevE.89.042811}
  {\path{doi:10.1103/PhysRevE.89.042811}}.

\bibitem{Lee2016}
D.~Lee, S.~Choi, M.~Stippinger, J.~Kert{\'{e}}sz, B.~Kahng, {Hybrid phase
  transition into an absorbing state: Percolation and avalanches}, Phys. Rev. E
  93~(4) (2016) 042109.
\newblock \href {https://doi.org/10.1103/PhysRevE.93.042109}
  {\path{doi:10.1103/PhysRevE.93.042109}}.

\bibitem{Zhou2014}
D.~Zhou, A.~Bashan, R.~Cohen, Y.~Berezin, N.~Shnerb, S.~Havlin, {Simultaneous
  first- and second-order percolation transitions in interdependent networks},
  Phys. Rev. E 90~(1) (2014) 012803.
\newblock \href {https://doi.org/10.1103/PhysRevE.90.012803}
  {\path{doi:10.1103/PhysRevE.90.012803}}.

\bibitem{Baxter2014}
G.~J. Baxter, S.~N. Dorogovtsev, J.~F.~F. Mendes, D.~Cellai, Weak percolation
  on multiplex networks, Phys. Rev. E 89~(4) (2014) 042801.

\bibitem{Hackett2016}
A.~Hackett, D.~Cellai, S.~G\'omez, A.~Arenas, J.~P. Gleeson, Bond percolation
  on multiplex networks, Phys. Rev. X 6~(2) (2016) 021002.

\bibitem{Son2011}
S.-W. Son, P.~Grassberger, M.~Paczuski, {Percolation Transitions Are Not Always
  Sharpened by Making Networks Interdependent}, Phys. Rev. Lett. 107~(19)
  (2011) 195702.
\newblock \href {https://doi.org/10.1103/PhysRevLett.107.195702}
  {\path{doi:10.1103/PhysRevLett.107.195702}}.

\bibitem{Li2012}
W.~Li, A.~Bashan, S.~V. Buldyrev, H.~E. Stanley, S.~Havlin, Cascading failures
  in interdependent lattice networks: The critical role of the length of
  dependency links, Phys. Rev. Lett. 108~(22) (2012) 228702.

\bibitem{Bashan2013}
A.~Bashan, Y.~Berezin, S.~V. Buldyrev, S.~Havlin, The extreme vulnerability of
  interdependent spatially embedded networks, Nat. Phys. 9~(10) (2013)
  667--672.

\bibitem{Berezin2013}
Y.~Berezin, A.~Bashan, S.~Havlin, {Comment on ``Percolation Transitions Are Not
  Always Sharpened by Making Networks Interdependent''}, Phys. Rev. Lett.
  111~(18) (2013) 189601.
\newblock \href {https://doi.org/10.1103/PhysRevLett.111.189601}
  {\path{doi:10.1103/PhysRevLett.111.189601}}.

\bibitem{Son2013}
S.-W. Son, P.~Grassberger, M.~Paczuski, {Son, Grassberger, and Paczuski
  Reply:}, Phys. Rev. Lett. 111~(18) (2013) 189602.
\newblock \href {https://doi.org/10.1103/PhysRevLett.111.189602}
  {\path{doi:10.1103/PhysRevLett.111.189602}}.

\bibitem{Grassberger2015}
P.~Grassberger, {Percolation transitions in the survival of interdependent
  agents on multiplex networks, catastrophic cascades, and solid-on-solid
  surface growth}, Phys. Rev. E 91~(6) (2015) 062806.
\newblock \href {https://doi.org/10.1103/PhysRevE.91.062806}
  {\path{doi:10.1103/PhysRevE.91.062806}}.

\bibitem{Jovanovic1994}
B.~Jovanovi{\'c}, S.~V. Buldyrev, S.~Havlin, H.~E. Stanley, Punctuated
  equilibrium and ``history-dependent'' percolation, Phys. Rev. E 50 (1994)
  R2403--R2406.

\bibitem{Hu2020}
M.~Hu, Y.~Sun, D.~Wang, J.-P. Lv, Y.~Deng, History-dependent percolation in two
  dimensions, Phys. Rev. E 102 (2020) 052121.
\newblock \href {https://doi.org/10.1103/PhysRevE.102.052121}
  {\path{doi:10.1103/PhysRevE.102.052121}}.

\bibitem{Li2020}
M.~Li, L.~L{\"u}, Y.~Deng, M.-B. Hu, H.~Wang, M.~Medo, H.~E. Stanley,
  History-dependent percolation on multiplex networks, National Science Review
  7~(8) (2020) 1296--1305.

\bibitem{Cai1990}
Z.-X. Cai, S.~D. Mahanti, S.~A. Solin, T.~J. Pinnavaia, Dual percolation
  threshold in two-dimensional microporous media, Phys. Rev. B 42 (1990)
  6636--6641.

\bibitem{Hansen1993}
A.~Hansen, E.~L. Hinrichsen, D.~Stauffer, {Percolation in layered media --- A
  conductivity approach}, Transport in Porous Media 11~(1) (1993) 45--52.

\bibitem{Kasteleyn1969}
P.~Kasteleyn, C.~Fortuin, Phase transitions in lattice systems with random
  local properties, Physical Society of Japan Journal Supplement 26 (1969) 11.

\bibitem{Reynolds1977}
P.~J. Reynolds, H.~E. Stanley, W.~Klein, Ghost fields, pair connectedness, and
  scaling: exact results in one-dimensional percolation, Journal of Physics A:
  Mathematical and General 10~(11) (1977) L203.

\bibitem{Stauffer2018}
D.~Stauffer, A.~Aharony, Introduction to percolation theory, Taylor \& Francis,
  2018.

\bibitem{Fortuin1972}
C.~M. Fortuin, {On the random-cluster model. III. The simple random-cluster
  model}, Physica 59~(4) (1972) 545--570.
\newblock \href {https://doi.org/10.1016/0031-8914(72)90087-0}
  {\path{doi:10.1016/0031-8914(72)90087-0}}.

\bibitem{Chen2013}
L.~Chen, F.~Ghanbarnejad, W.~Cai, P.~Grassberger, Outbreaks of coinfections:
  The critical role of cooperativity, Europhysics Letters 104~(5) (2013) 50001.
\newblock \href {https://doi.org/10.1209/0295-5075/104/50001}
  {\path{doi:10.1209/0295-5075/104/50001}}.

\bibitem{Min2020}
B.~Min, C.~Castellano, Message-passing theory for cooperative epidemics, Chaos:
  An Interdisciplinary Journal of Nonlinear Science 30~(2) (2020) 023131.
\newblock \href {https://doi.org/10.1063/1.5140813}
  {\path{doi:10.1063/1.5140813}}.

\bibitem{Lee2014}
K.-M. Lee, C.~D. Brummitt, K.-I. Goh, Threshold cascades with response
  heterogeneity in multiplex networks, Phys. Rev. E 90 (2014) 062816.
\newblock \href {https://doi.org/10.1103/PhysRevE.90.062816}
  {\path{doi:10.1103/PhysRevE.90.062816}}.

\bibitem{Min2018}
B.~Min, M.~San~Miguel, Competition and dual users in complex contagion
  processes, Scientific Reports 8~(1) (2018) 14580.
\newblock \href {https://doi.org/10.1038/s41598-018-32643-4}
  {\path{doi:10.1038/s41598-018-32643-4}}.

\bibitem{Pocock2012}
M.~J.~O. Pocock, D.~M. Evans, J.~Memmott, The robustness and restoration of a
  network of ecological networks, Science 335~(6071) (2012) 973--977.
\newblock \href {https://doi.org/10.1126/science.1214915}
  {\path{doi:10.1126/science.1214915}}.

\end{thebibliography}
\bibliographystyle{elsarticle-num}

\end{document}